\documentclass[preprint,showpacs,prb]{revtex4}%
\usepackage{amsfonts}
\usepackage{amsmath}
\usepackage{amssymb}
\usepackage{graphicx}%
\begin{document}
\title{Electric-field switching of exciton spin splitting in coupled quantum dots}
\author{Xiaojing Li}
\author{Kai Chang}
\email{kchang@red.semi.ac.cn}
\affiliation{SKLSM, Institute of Semiconductors, Chinese Academy of Sciences, P. O. Box
912, Beijing 100083, China}

\pacs{68.65.Hb, 61.72.sd, 81.40.Rs}

\begin{abstract}
We investigate theoretically the spin splitting of the exciton states in
semiconductor coupled quantum dots (CQDs) containing a single magnetic ion. We
find that the spin splitting can be switched on/off in the CQDs \textit{via}
the \textit{sp-d }exchange interaction using the electric field. An
interesting bright-to-dark exciton transition can be found and significantly
affects the photoluminescence spectrum. This phenonmenon is induced by the
transition of the ground exciton state, arising from the hole mixing effect,
between the bonding and antibonding states.

\end{abstract}
\maketitle

All-electrical control of electron spin is a central goal in the field of
spintronics and quantum information processing. There have been many proposals
concerning the experimental realization of all-electrical spintronic devices
and qubits in solid state systems in recent years. The electron spin in
quantum dots (QD) is a promising candidate for the qubit\cite{Divi2} due to
very long spin decoherent time and the feasibility of large-scale integration.
There are several different schemes to control carrier spin in semiconductor
QDs, e.g., the circular-polarized optical excitation, the spin-orbit
interaction, and the \textit{sp-d} exchange interaction. The \textit{sp-d}
exchange interaction between the carriers and the magnetic ions in\ a QD leads
to giant Zeeman splitting and could be an important testing ground for the
realization of a solid-state qubit.\cite{Kim,Kossut,Kai} The Photoluminesence
(PL) spectrum of CdTe QDs doped with a single Mn$^{2+}$ ion have demonstrated
the effect of the \textit{sp-d} exchange interaction on the interband
transition.\cite{Beso,Maksimov,Xin} This provides a unique flexibility to
tailor the spin splitting of carriers and optical property utilizing external
electric fields. However, a strong electric field is required to tune the spin
splitting of the exciton in a QD containing a single magnetic ion since the
strong confinement of carriers in a single QD prohibit the spatial separation
of the electron and hole.\cite{Beso,LiXJ} Is it possible to control the spin
states of exciton easily\ utilizing weak electric field in the QD system?

Recently, it was demonstrated that the orbital states of carriers in coupled
QDs (CQDs) can be tuned by an external electric field in vertically- and
laterally-coupled QDs, respectively.\cite{Ortner,Peeters} An interesting
field-induced dissociation of exciton was observed in Photoluminesence
experiments.\cite{Krenner,Scheibner} In this work, we consider the CQDs doped
with a single Mn$^{2+}$ ion and find that the spin states of the exciton in
such CQDs can be easily controlled using electric fields. The spin splitting
of the exciton in the CQDs exhibits significant asymmetry with respect to the
directions of the electric fields and a switching behavior for the weak
coupling case. An interesting bright-to-dark exciton transition arising from
the bonding-antibonding hole state transition can be seen by adjusting the
parameters, e.g., the spatial separation $d$ between the two QDs.

The CQDs structure is schematically shown in Fig. \ref{fig:1}. The electron
and the hole are confined by a square potential well in the z axis,
\begin{equation}
V_{\perp}^{e,h}(z)=\left\{
\begin{array}
[c]{c}%
0,[-(d/2+d_{1}),-d/2]\\
0,[d/2,(d/2+d_{2})]\\
\Delta V^{e,h},otherwise
\end{array}
\right.  ,\label{z_potential}%
\end{equation}
and laterally by a parabolic potential $V_{\parallel}^{e,h}(\rho
)=m_{e,h}\omega_{e,h}^{2}\rho^{2}/2$, where $m_{e}$ and $m_{h}$ are the
effective mass of the electron and heavy hole, respectively. The electric
field is applied perpendicular to the CQDs plane. The Hamiltonian of the
system is $H=H_{e}+H_{h}+H_{s-d}+H_{p-d}+H_{e-h}+V_{Coul}$, where $H_{e}$
($H_{h}$) is the electron (hole) Hamiltonian, $H_{s-d}$ ($H_{p-d}$) is the
\textit{s-d} (\textit{p-d}) exchange interaction between the electron spin
$\mathbf{s}$ (hole spin $\mathbf{j}$) located at $\mathbf{r}_{e}$
($\mathbf{r}_{h}$) and the Mn$^{2+}$ spin $\mathbf{S}$ located at
$\mathbf{r}_{Mn}$, and $H_{e-h}$ is the short-range exchange interaction
between the electron and hole. $V_{Coul}=-e^{2}/4\pi\varepsilon_{0}%
\varepsilon|\mathbf{r}_{e}-\mathbf{r}_{h}|$ is the Coulomb interaction between
the electron and hole, where $\varepsilon$($\varepsilon_{0}$) is the
dielectric constant of the material (vacuum), and $e$ is the charge of the
electron. The hole Hamiltonian is\emph{\ }$H_{h}=H_{LK}-eEz$\emph{, }where
$H_{LK}$ is the four-band Luttinger-Kohn Hamiltonian including the heavy hole
and light hole bands.\cite{Kai} The relevant material parameters can be found
in Ref. [6]. The eigenstate of the exciton $\Psi^{eh}(\mathbf{r}%
_{e},\mathbf{r}_{h})$ is expanded in the basis set constructed by the direct
product of the eigenstates of the electron, the hole, and the magnetic
ion.\cite{LiXJ}

Fig. \ref{fig:2} (a) displays the exciton energy spectrum as a function of
electric field in a CdTe CQDs with strong interdot coupling ($d=2nm$). We
magnify the energy spectrum at weak electric fields (see the inset in Fig.
\ref{fig:2} (a)) and find a strong asymmetric spin splitting with respect to
the directions of the applied electric fields. The electron and hole
distribute equally in the upper and lower QDs in the absence of the electric
field. An external electric field pushes the electron and hole in opposite
directions and consequently leads to the localization of electron and hole in
different QDs. Compared to a single QD tuned by the external field, the
wavefunction of carriers in such CQDs is much more easily manipulated
electrically due to the weakening of the quantum confinement along the z axis.
Since the strength of the $p-d$ exchange interaction is about four times
larger than that of the $s-d$ interaction, when the hole is pushed into the
upper QD that contains a single magnetic ion (see Fig. \ref{fig:1}), the spin
splitting becomes larger. The spin splitting becomes very small when the
electron localizes in the upper QD. The twelve highest exciton energy states
in the spectrum consist of antibonding hole-states (the first-excited states
of the spin-down hole $|3/2,-3/2>$) and the electron ground state
$|1/2,\pm1/2>$. Meanwhile, the hole states of the twelve lowest exciton energy
levels are composed of the bonding spin-down hole states $|3/2,-3/2>$ and the
electron ground state $|1/2,\pm1/2>$. The energy difference between the
bonding and antibonding exciton states can be seen in Fig. \ref{fig:2} (b).
The electron and hole states in the twelve lowest exciton states are both
bonding states, but we can only see six bright lines for the $\sigma^{+}$
excitation (the other six exciton levels ($J=\pm2$) are dark states). However,
the hole states of the twelve highest exciton energy levels are the
antibonding states, therefore they are dark states at the small electric field
in the electro-PL spectrum. As the electric field increases, we find the dark
antibonding exciton states become bright, arising from the mixing of the
bonding-antibonding hole states. There is an energy gap between the bonding
and antibonding exciton states because of the interdot tunneling coupling
between the two QDs. The bright $\sigma^{\pm}$ $(\pm\frac{1}{2},\mp\frac{3}%
{2},S_{z})$ and dark $\pm2$ $(\pm\frac{1}{2},\pm\frac{3}{2},S_{z})$ exciton
states also split because of $s-d$ interaction. Spin splitting of the bonding
exciton states increases/decreases oppositely to that of the antibonding
exciton states as the electric field varies. In such CQDs, we can realize a
switching behavior for the spin splitting utilizing weak electric fields.

For comparision with the CQDs with the strong interdot coupling, we calculate
the CQDs for weak interdot coupling case($d=4nm$). The exciton energy spectrum
in Fig. \ref{fig:3}(a) shows that the energy gap between the bonding and
antibonding exciton states decreases because of the weakening of the interdot
tunneling coupling. The energy spectrum resembles that of the strong coupling
case, while the electro-PL spectrum is very different (see Fig. \ref{fig:3}%
(b)). This is because the twelve lowest exciton states are no longer the
bright states, since the hole component of the exciton states shows an
antibonding feature. In contrast, the highest twelve exciton states become
bright, i.e., the bonding hole states in the exciton states. We plot the hole
energy as a function of the spatial separation $d$ between the QDs in Fig.
\ref{fig:4}(b). A crossover between the bonding and antibonding hole states
takes place around $d=2.2nm$ (see the arrow in Fig. \ref{fig:4} (b)). The
crossover can also be seen in Fig. \ref{fig:4}(c), which plots the overlap
factor between the electron and hole states as a function of the distance $d$
(see the black lines in \ref{fig:4}(c)). There is no crossover between the
ground and first-excited hole states, i.e., the bonding and antibonding hole
states, without coupling of heavy hole (\textit{hh}) and light hole
(\textit{lh}) mixing (see the dotted lines in Fig.\ref{fig:4}(b)). This
behavior can be understood from a four-level model that is schematically shown
in Fig. \ref{fig:4}(a). In the Luttinger-Kohn Hamiltonian, the off-diagonal
element $R\propto k_{\Vert}k_{z}$ induces the coupling between the HH($L=0$)
and LH($L=1$) states (see Fig. 4(a)). The lowest two levels $\lambda
_{i}=E_{01}^{i}/2\pm\sqrt{(E_{01}^{i})^{2}+4\Delta_{i}^{2}}/2$ ($i=a,b)$,
where $E_{01}^{a}=E_{0}^{a}+E_{1}^{b}(E_{01}^{b}=E_{0}^{b}+E_{1}^{a})$ is the
sum of the energies of the two original coupled levels and $\Delta_{i}$ is the
coupling term. The competition between $E_{01}^{i}$ and $\Delta_{i}$ could
lead to the bonding and antibonding transition of the ground state. Increasing
the distance $d$ results in the change of the coupling between the HH
(bonding) and LH (antibonding) states $\Delta_{i}$. As a result, the energy of
antibonding exciton states could be lower than bonding exciton states.
Therefore, the the twelve lowest exciton states become dark while the twelve
highest states are bright in the electro-PL spectrum. This feature can also be
understood from the overlap factor of the bonding and antibonding exciton
states as a\ function of electric field (see the red lines in Fig. \ref{fig:4}
(c)). We find that the overlap factor of the antibonding (bonding) exciton
states increase as the electric field increases (decreases), resulting in the
dark-to-bright transition of the ground exciton states. Because the exciton
energies of bonding and antibonding states are very close to each other, it is
hard to distinguish them from the PL\ spectrum. The electro-PL spectrum could
help us to distinguish them from the intensity of the PL\ peaks\ due to the
different energy dependences of the bonding and antibonding exciton states on
electric fields. (see Figs. 2 and 3)

Finally, we plot the spin splitting as function of the electric field and the
distance $d$ between the CQDs in Fig. \ref{fig:5}. The spin splitting is
symmetric with respect to the directions of electric fields when the CQDs are
strongly coupled. The weak electric fields only result in a negligiable small
spin splitting since the strong confinement of carriers prohibits the spatial
separation of the electron and hole. When the distance $d$ increases, there is
different behavior of the spin splitting at weak electric fields. The spin
splitting becomes strongly asymmetric with respect to the directions of the
electric fields and shows a switching feature for the opposite directions of
the electric fields. We should point out that the position of the magnetic ion
in the QD affects heavily the exciton spin splitting, and determines the
electric field corresponding to the largest spin splitting, but it will not
change the switching behavior in CQDs containing a single magnetic ion.

In summary, we investigated theoretically the energy spectrum and electro-PL
spectrum of the CQDs containing a single magnetic ion. For the CQDs with
strong interdot coupling, the spin splitting is asymmetric with respect to the
directions of the electric fields and\ can be switched on/off using weak
electric fields. For the weak coupling case, we find that the hole mixing
effect leads to the crossover between the bonding and antibonding hole states,
consequently resulting in the bright-to-dark transition of the ground exciton
states. Our theoretical results could be useful for the designing fresh types
of all-electrical spintronic devices.

\begin{acknowledgments}
This work was supported by the NSFC Grant No. 60525405.
\end{acknowledgments}

\newpage


\begin{thebibliography}{99}                                                                                               %


\bibitem {Divi2}D. P. DiVincenzo, D. Bacon, J. Kempe, G. Burkard. K. B.
Whaley, Nature (London) \textbf{408}, 339 (2000).

\bibitem {Kim}C. S. Kim, M. Kim, S. Lee, J. Kossut, J. K. Furdyna, and M.
Dobrowlska, J. Cryst. Growth \textbf{214}, 395 (2000).

\bibitem {Kossut}J. Kossut, I. Yamakawa., A. Nakamura and S. Takeyama, Appl.
Phys. Lett. \textbf{79}, 1789 (2001).

\bibitem {Kai}Kai Chang, J. B. Xia, F. M. Peeters, Appl. Phys. Lett.
\textbf{82}, 2661 (2003); Kai Chang, S. S. Li, J. B. Xia, and F. M. Peeters,
Phys. Rev. B \textbf{69}, 235203 (2004).

\bibitem {Beso}Y. L\'{e}ger, L. Besombes, J. Fern\'{a}ndez-Rossier, L.
Maingault, and H. Mariette, Phys. Rev. Lett. \textbf{97}, 107401 (2006); L.
Besombes, Y. Leger, L. Maingault, and H. Mariette, J. Appl. Phys.
\textbf{101}, 081713 (2007).

\bibitem {Maksimov}A. A. Maksimov, G. Bacher, A. MacDonald, V. D. Kulakovskii,
A. Forchel, C. R. Becker, G. Landwehr, and L. Molenkamp, Phys. Rev. B
\textbf{62}, R7767 (2000).

\bibitem {Xin}S. H. Xin, P. D. Wang, A. Yin, C. Kim, M. Dobrowolska, J. L.
Merz, and J. K. Furdyna, Appl. Phys. Lett. \textbf{69}, 2884 (1996).

\bibitem {LiXJ}X. J. Li, and Kai Chang, Appl. Phys. Lett. \textbf{92}, 071116(2008).

\bibitem {Ortner}G. Ortner, M. Bayer, Y. Lyanda-Geller, T. L. Reinecke, A.
Kress, J. P. Reithmaier, and A. Forchel, Phys. Rev. Lett. \textbf{94}, 157401 (2005).

\bibitem {Peeters}B. Szafran and F. M. Peeters, Phys. Rev. B \textbf{76},
195442 (2007).

\bibitem {Krenner}H. J. Krenner, M, Sabathil, E. C. Clark, A. Kress, D. Schuh,
M. Bichler, G. Abstreiter, and J. J. Finley, Phys. Rev. Lett. \textbf{94},
057402 (2005).

\bibitem {Scheibner}M. Scheibner, M. F. Doty, I. V. Ponomarev, A. S. Bracker,
E. A. Stinaff, V. L. Korenev, T. L. Reinecke, and D. Gammon, Phys. Rev. B
\textbf{75}, 245318 (2007).
\end{thebibliography}

\newpage
\begin{figure}[htbp]
\caption{Schematic diagram
of a CQD containing a single magnetic ion.}%
\label{fig:1}%
\end{figure}

\begin{figure}[htbp]
\caption{The exciton energy spectrum as a function of electric field
(the upper panel) and the electro-PL spectrum (the lower panel) in
CQD. $d_{1}=d_{2}=2.4nm$, the radius
$l_{e}=l_{h}=5nm$, the distance between the two QDs $d=2nm$.}%
\label{fig:2}%
\end{figure}

\begin{figure}[htbp]
\caption{The same as Fig. 2,
but for the CQD with larger spatial separation between the two QDs, $d=4nm$.}%
\label{fig:3}%
\end{figure}

\begin{figure}[htbp]
\caption{(Color online)(a) the schematic diagram of the four-level
model. (b) The hole energy as a function of distance $d$ between the
two QDs. The solid lines denote the hole energies in CQDs with the
HH-LH coupling; the dotted lines denote that without the HH-LH
coupling. (c) the overlap factor between the electron ground state
and the hole ground states (the solid lines) and first-excited hole
states (the dotted lines) as function of the distance $d$ between
the two QDs(the
black lines) and the electric fields (the red lines) for $d=4nm$.}%
\label{fig:4}%
\end{figure}

\begin{figure}[htbp]
\caption{The contour plot of the spin splitting of exciton in a CQD
as function of the distances
$d$ and external electric fields.}%
\label{fig:5}%
\end{figure}

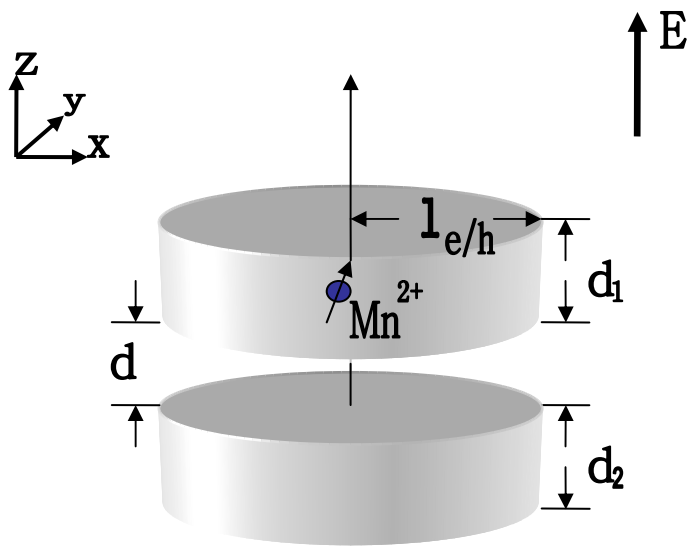
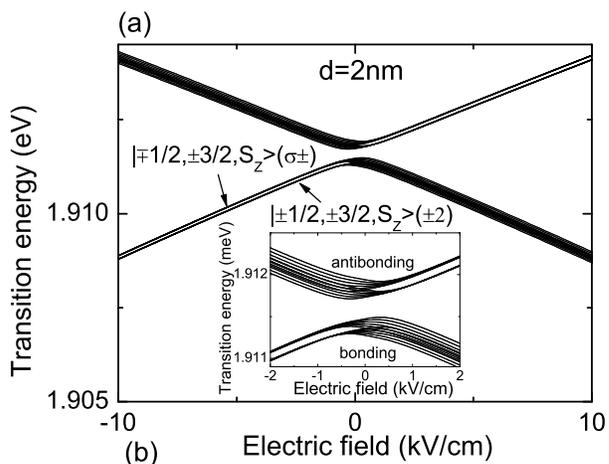
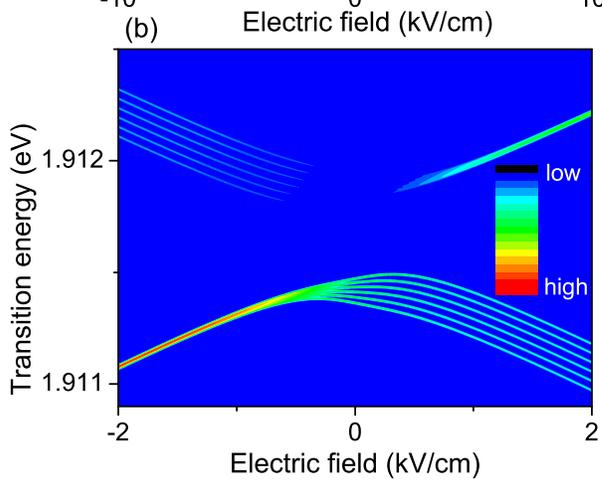
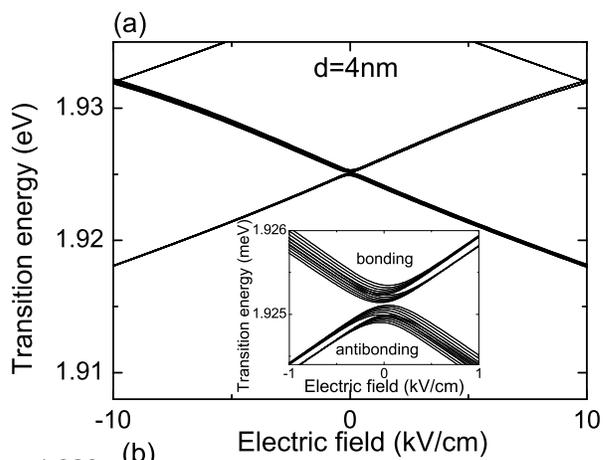
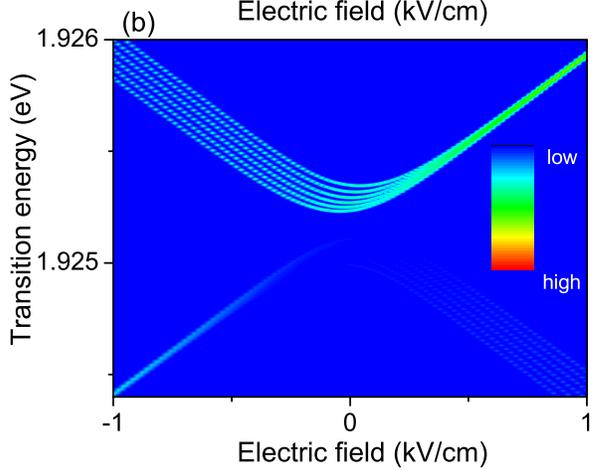
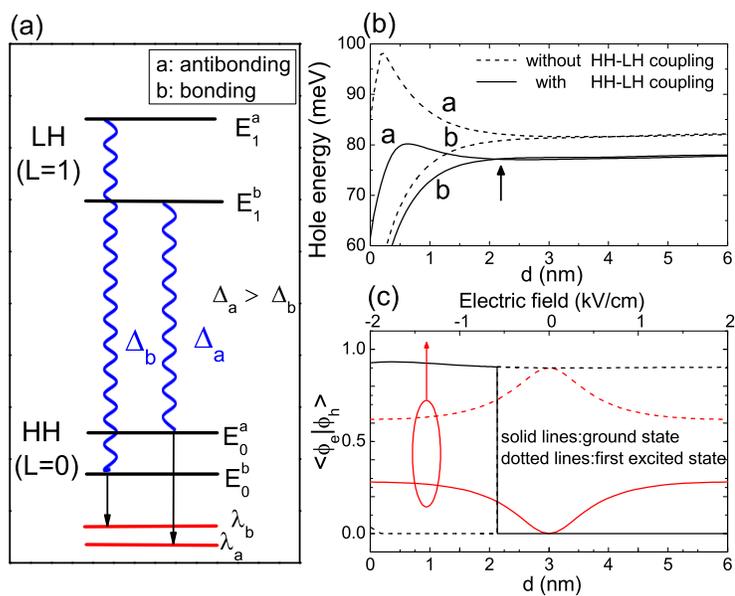
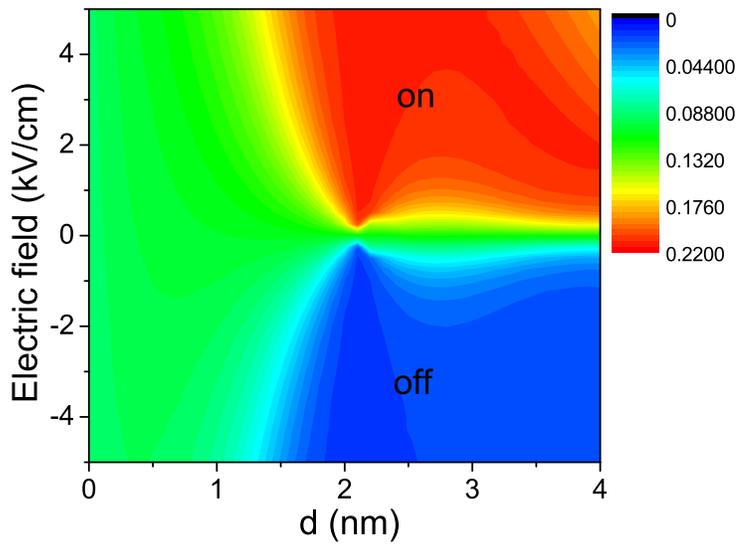
\end{document}